\documentclass[final,3p,times,twocolumn]{elsarticle}
 \biboptions{comma,sort&compress}

\usepackage{amsmath}

\def\nin{\noindent}
\def\beq{\begin{equation}}
\def\eeq{\end{equation}}
\def\bea{\begin{eqnarray}}
\def\eea{\end{eqnarray}}

\journal{Nuc. Phys. (Proc. Suppl.)}

\begin{document}

\begin{frontmatter}



\title{
\vspace*{-1cm}
{ \hspace*{13cm}{\bf \small FTUAM-14-26}  }
\\
\vspace*{-0.2cm}
{ \hspace*{13cm}{\bf \small IFT-UAM/CSIC-14-064}   }
\\
\vspace*{+1cm}
{\large Assessing the accuracy of the Brodsky-Lepage prediction}     }

 \author[label1]{Pablo Roig\corref{cor1}}
  \address[label1]{Instituto de F\'{\i}sica, Universidad Nacional Aut\'onoma de M\'exico,
Apartado Postal 20-364, 01000 M\'exico D.F., M\'exico.}
\cortext[cor1]{Speaker}
\ead{pabloroig@fisica.unam.mx}

 \author[label2]{Juan Jos\'e Sanz-Cillero}
  \address[label2]{Departamento de F\'{\i}sica Te\'orica
and Instituto de F\'{\i}sica Te\'orica, IFT-UAM/CSIC, Universidad Aut\'onoma de Madrid, Cantoblanco, E-28049 Madrid, Spain.}
\ead{juanj.sanz@uam.es}


\begin{abstract}
\noindent
We show that a consistent set of short-distance conditions on the odd-intrinsic parity resonance chiral Lagrangian couplings can be derived in the large-$N_C$ limit and
the single resonance approximation. These constraints are satisfied both by the VVP Green function and all its associated form-factors. Moreover, they are in agreement with
earlier results found in the normal parity sector, which further supports the overall consistency of the approach. Here we take a new perspective and examine the consequences
of these high-energy relations on the precise coefficient of the $1/Q^2$-term of meson form-factors, which rules their asymptotic behavior. In particular, we show that the
coefficient derived using our consistent set of short-distance constraints is in remarkable accordance with the Brodsky-Lepage prediction,
being alternative results for these coefficients disfavored.
\end{abstract}


\end{frontmatter}


\section{Introduction: short-distance QCD constraints and their role in the resonance region}
\label{Intro}
\nin
Non-perturbative QCD is a rather complicated subject, whose solution remains elusive. In particular, analytic approaches to the dynamics of the lowest-lying light-flavored
resonances have proved to be way more difficult than to the lightest pseudoscalar mesons, for which an effective field theory dual to QCD at low energies (Chiral
Perturbation Theory, $\chi PT$) exists \cite{ChPT}. The main reason behind is the difficulty to find a suitable expansion parameter in the resonance region, where the
$\chi PT$ one, in powers of momenta and light meson masses over the chiral symmetry breaking scale, is no longer small and thus non-applicable.\\
A formal expansion for a large number of colours \cite{Nc} in powers of $1/N_C$ succeeds to explain the most salient features of meson phenomenology, both at the fundamental
and at the effective field theory levels \cite{Nc_eff} with a typical accuracy that turns out to be much better than the naive $1/3$ estimate.\\
A concrete realization of these ideas are resonance chiral Lagrangians ($R\chi L$) \cite{RChT}, which are built adding to the lowest order $\chi PT$ Lagrangian (which corresponds
to the realization of the chiral anomaly in $\chi PT$ \cite{WZW} in the odd-intrinsic parity sector) operators with an increasing number of resonance fields and chiral operators
appearing already in the $\chi PT$ framework. Chiral symmetry for the latter and $SU(3)$ symmetry for the former determine the allowed operators, leaving however their
couplings unrestricted.\\
It is useful to recall at this point that not only the low-energy limit of meson dynamics is known. On the other end, the Operator Product Expansion (OPE) of QCD \cite{OPE}
allows to derive its short-distance behavior as well. $R\chi L$ are thus thought to provide a bridge between these two known regimes. This interpolation is more reliable for
the so-called order parameters of chiral symmetry breaking (the perturbative contribution to them vanishes to all orders), where more precisely we will be considering
Green functions and related form-factors. In some sense, this procedure implements at the Lagrangian level Pad\'e and related rational approximants studied in previous works~\cite{Pades}.
Thus, one should be aware of the uncertainties and caveats raised therein when the infinite large-$N_C$ spectrum is truncated and one just keeps the lightest resonances.\\
A well-known example that illustrates all the above easily comes from the Weinberg sum rules \cite{WSR}. According to them, QCD predicts that the two-point vector correlator
between left- and right-handed quark fields vanishes at least as fast as $1/Q^4$ when the squared Euclidean momentum, $Q^2$, goes to infinity. The corresponding evaluation within $R\chi L$
\cite{RChT} restricted to the lightest meson multiplets (single resonance approximation \footnote{This single resonance approximation will be considered unless otherwise
specified and subleading $1/N_C$ corrections will be neglected.}) yields, in the chiral limit,
\begin{equation}\label{WSR}
 \Pi_{LR}(Q^2)\,=\,\frac{F^2}{Q^2}\,+\,\frac{F_V^2}{M_V^2-Q^2}\,-\,\frac{F_A^2}{M_A^2-Q^2}\,,
\end{equation}
and the high-energy limit mentioned above imposes
\begin{equation}\label{WSRrestrictions}
 F_V^2\,-\,F_A^2\,=\,F^2\,,\quad M_V^2\, F_V^2\,-\,M_A^2\, F_A^2\,=\,0\,,
\end{equation}
giving the \textit{a priori} free coupling of the (axial-)vector resonance to the (axial-)vector current, ($F_A$) $F_V$ in terms of the corresponding masses and the
pion decay constant, $F$.\\
We will finish this introduction by recalling that there are two ways of approaching the asymptotic behavior of QCD in order to demand the correct short-distance limit:
\begin{itemize}
 \item The high-energy QCD limit of a given Green function ($VVP$ in what follows) is computed using the OPE and this behavior is
required for the corresponding computation using a $R \chi L$.
 \item The QCD one-loop result for the
imaginary part of the (axial-)vector--(axial-)vector correlator \cite{Floratos:1978jb} predicts that, at high energies, it must go to a definite constant value. By the
optical theorem, this absorptive contribution can be equivalently obtained through a sum over the infinite number of possible intermediate on-shell cuts,
given by the corresponding form-factors. A plausible assumption for recovering a constant through the infinite sum is that
the contribution from each channel vanishes as $1/Q^2$ asymptotically, reproducing the celebrated Brodsky-Lepage prediction \cite{BL}. It must be noted that the latter
also fixes the corresponding coefficient of the $1/Q^2$  term.
\end{itemize}
A seeming discrepancy between some of the constraints in the anomaly sector  was found depending on whether one used one or the other procedure sketched
above~\cite{RuizFemenia:2003hm, Dumm:2009kj}.
This situation led us to a reanalysis of the compatibility of
both approaches: our results \cite{Us} showed that they were indeed equivalent, as it was also shown in the normal parity
case \cite{even-intrinsic}. Specifically, we show the existence of a consistent set of short-distance constraints on the anomalous resonance couplings within the single
resonance approximation and in the $N_C\to\infty$ limit~\footnote{The short-distance study of more complicated
form-factors and/or Green functions may disagree with these relations \cite{VFRO}.}. Representing the spin-one fields with antisymmetric tensors turns out to be crucial for
this result.\\
New results appear in Sec.~\ref{New}, where we use ultraviolet constraints to test the precise coefficient
of the leading $1/Q^2$ term in the high-energy
expansion of the neutral and charged one-pion vector form-factors, related to the $\pi^0\to\gamma\gamma^*$ and $\tau^-\to \nu_\tau \pi^-\gamma$ decays, respectively.
Our results are in much better agreement with the Brodsky-Lepage coefficient than with other predictions.
\section{The case of the anomalous sector of QCD}
\label{Odd}
\nin
Ref.~\cite{RChT} showed that, in the even-intrinsic parity sector, it was possible to match the QCD short-distance behavior (for a selected set of Green functions and form
factors) with a resonance chiral Lagrangian. Noteworthy, it showed that representing spin-one resonances in the antisymmetric tensor formalism, local operators of the next-to-leading
order (NLO) $\chi PT$ Lagrangian were not needed for this matching. These operators without resonance fields were nevertheless needed if one chose the four-vector Proca
representation for the spin-one mesons. This suggested the convenience of the antisymmetric tensor formalism, as it provided a simpler $R\chi L$.\\
Conversely, Refs.~\cite{VFRO, Knecht:2001xc} showed that the anomalous sector was more complicated. Specifically, Ref.~\cite{Knecht:2001xc} concluded that the four-vector
representation could not match the OPE results even including higher-order $\chi PT$ operators, an important result which casted serious doubts on the usefulness of the
resonance Lagrangians in the anomalous sector.\\
This fact motivated the study of Ref.~\cite{RuizFemenia:2003hm} showing that the matching was indeed possible for the VVP Green function using the antisymmetric tensor
representation and definite predictions were obtained for several odd-intrinsic parity resonance couplings. Although the puzzle seemed to be solved, the analysis of
short-distance QCD constraints on the vector form-factors of $\tau^-\to (KK\pi)^-\nu_\tau$ decays \cite{Dumm:2009kj} did not agree those in Ref.~\cite{RuizFemenia:2003hm}.
The latter study was extended by Ref.~\cite{Kampf:2011ty} adding the lightest pseudoscalar resonances as explicit degrees of freedom and proving
the saturation of the chiral constants at NLO in the anomalous sector.\\
The aim of our work \cite{Us} was to clarify if the inclusion of pseudoscalar resonances as active degrees of freedom for the VVP calculation \cite{Kampf:2011ty} was able to
solve the previously existing discrepancies in the high-energy constraints on the $R\chi L$ couplings stemming from all the available form-factor and $VVP$ Green function analyses at large $N_C$.\\
The detailed discussion can be found in Ref.~\cite{Us}, where it is seen that there is a unique set of consistent high-energy constraints in the odd intrinsic-parity sector
\footnote{We only write these constraints in terms of the $c_i$ and $d_j$ couplings introduced in Ref.~\cite{RuizFemenia:2003hm}. The corresponding five first equations in
terms of the $\kappa_k$ couplings in Ref.~\cite{Kampf:2011ty} and the dictionary between both operator bases is provided in our letter \cite{Us}.}
\begin{eqnarray}
& & 4 \,c_3 \,+\, c_1  = \quad  0\,,
\nonumber
\\[3mm]
& & c_1 \, - \, c_2 \, + \, c_5 \,  = \quad  0 \, ,
\nonumber
\\[3mm]
& & c_5 \, - \, c_6 \,  =
\quad  \frac{N_C \,M_V}{64 \,\sqrt{2}\, \pi^2\,F_V} \, ,
\nonumber
\\[3mm]
& &  d_1 \, + \, 8 \,d_2  = \quad
\frac{F^2}{8\,F_V^2} - \frac{N_C \,M_V^2}{64\, \pi^2\, F_V^2}\,,
\nonumber
\\[3mm]
& & d_3 =\quad  -\, \frac{N_C}{64\pi^2} \frac{M_V^2}{F_V^2} \,,
\nonumber
\\[3mm]
& & 1 \, + \, \frac{32 \,\sqrt{2} \, F_V \,d_m\, \kappa_3^{PV}}{F^2} = \quad  0\,,
\nonumber
\\[3mm]
& & F_V^2 = \quad  3\, F^2\,,
\label{eq.consistent-set-of-relations}
\end{eqnarray}
compatible for the VVP Green function~\cite{RuizFemenia:2003hm, Kampf:2011ty}
and a series of related odd intrinsic-parity amplitudes: the $\tau\to X^- \nu_\tau$
vector form-factors ($X^-=(KK\pi)^-$~\cite{Dumm:2009kj}, $\varphi^-\gamma$~\cite{Guo:2010dv},
$(\varphi V)^-$~\cite{Guo:2008}, with $\varphi=\pi,\,K$) and the $\pi^0\to\gamma^*\gamma^*$ transition form factor~\cite{Kampf:2011ty}.
The last relation in eqs.~(\ref{eq.consistent-set-of-relations}), which is also found in the normal parity sector, was not noticed in Ref.~\cite{Kampf:2011ty}. The previous
one, involving pseudoscalar resonance couplings \cite{Kampf:2011ty}, is also crucial for the overall consistency. Short-distance constraints from other anomalous processes
\cite{Others} agree with eqs.~(\ref{eq.consistent-set-of-relations}). Their (small) modification
upon the inclusion of heavier resonances \cite{Radpidec} and the (successful) description of related phenomenology is discussed in detail in our letter \cite{Us}.\\
We have applied eqs.~(\ref{eq.consistent-set-of-relations}) to study the lightest pseudoscalar ($\pi^0$, $\eta$, $\eta'$) transition form-factors and the corresponding
contribution to the hadronic light-­‐by-­‐light scattering piece of the muon $g-2$ \cite{Roig:2014uja} with the result $(10.47\pm0.54)\cdot10^{-10}$, which would
shrink the theoretical error on the muon anomaly, $a_\mu$, by $\sim15\%$ \cite{Knecht}. All uncertainties contributing to the above error are detailed in Ref.~\cite{Roig:2014uja}.
We note that the reduced uncertainty comes basically from the appearance of new data on the meson transition form-factors extending to larger energies and not from our
refined theoretical setting.\\
Checking the reliability of the resonance coupling determinations is specially important for guiding global phenomenological studies as those being worked out \cite{TAUOLAnew}
within the TAUOLA Monte Carlo generator \cite{TAUOLA}.
\subsection{Testing Brodsky-Lepage predictions}\label{New}
We consider now our consistent set of asymptotic constraints on the anomalous resonance
couplings (\ref{eq.consistent-set-of-relations}) from another perspective that sheds light on the coefficient of the leading $1/Q^2$-term in the asymptotic
high-energy expansion of meson form-factors. We will in particular consider three predictions, using light-cone perturbation theory \cite{BL}, the OPE \cite{M} and the
Bjorken-Johnson-Low theorem \cite{GL}.\\
The vector form factor $\mathcal{F}_{\pi\gamma}^V(Q^2)$, which enters in the $\tau^-\to \nu_\tau \pi^-\gamma$ decay,
was analyzed in Ref.~\cite{Guo:2010dv} requiring the vanishing of the $\mathcal{O}(Q^2)$ and $\mathcal{O}(Q^0)$
terms.
These constraints lead to the prediction
\begin{equation}
Q^2 \mathcal{F}_{\pi\gamma}^V(Q^2) \xrightarrow{Q^2\to\infty} M_V^2 N_C/(24 \pi^2 F)  \, ,
\end{equation}
to be compared to previous results for the leading term in the asymptotic expansion,
\begin{eqnarray}
\hspace*{-0.75cm}
 Q^2 \mathcal{F}_{\pi\gamma}^V(Q^2) \xrightarrow{Q^2\to\infty} F\,\,\, \text{\cite{BL}}\,,\,\,\,   \frac{2F}{3}\,\,\,\text{\cite{M}}\,,\,\,\,   \frac{F}{3}\,\,\,\text{\cite{GL}}\, .
\end{eqnarray}
For the inputs $M_V=760$~MeV and  $F=90$~MeV the deviations between our result and the latter are $\sim 10\%$, $\sim 35\%$ and $\sim 270\%$, respectively,
being the Brodsky-Lepage behavior \cite{BL} preferred.\\
A similar exercise can be made with the pion transition form factor with a real and a virtual photon,
$\mathcal{F}_{\pi^0\gamma\gamma^\star}(Q^2)$, in the chiral limit.
Demanding that this form-factor vanishes at high energies yields our R$\chi$L  prediction
\begin{equation}
Q^2 \mathcal{F}_{\pi^0\gamma\gamma^*}(Q^2) \xrightarrow{Q^2\to\infty} M_V^2 N_C/(12 \pi^2 F)  \, .
\end{equation}
Correspondingly, one has the previous results, 
\begin{equation}\label{piTFF_pred}
 Q^2 \mathcal{F}_{\pi^0\gamma\gamma^\star}(Q^2) \xrightarrow{Q^2\to\infty} 2F\,\,\, \text{\cite{BL}}\,,  \,\,\, \frac{4F}{3}\,\,\,\text{\cite{M}}\,, \,\,\, \frac{2F}{3}\,\,\,\text{\cite{GL}}\,,
\end{equation}
where the deviations between the latter and our R$\chi$L  outcome is the same as for $\mathcal{F}^V_{\pi\gamma}(Q^2)$.
\\
Therefore, closer agreement with the Brodsky-Lepage result~\cite{BL} is found
for both form-factors. These estimates justify the demand of $Q^2\mathcal{F}_{\pi\gamma}^V(Q^2)\sim F$ for large $Q^2$ required in Ref.~\cite{Guo:2010dv}.
We want to remark that the constraints from  these form-factors are fully compatible with the high-energy conditions
extracted  from the VVP Green function and the remaining related form-factors, provided in eq.~(\ref{eq.consistent-set-of-relations}).
\\
\section{Conclusions}
\nin
A consistent minimal set of short-distance constraints on the anomalous $R\chi L$ couplings which applies to the VVP Green function and related form-factors has been found
solving a seeming discrepancy in the literature. Interestingly, although eq.~(\ref{eq.consistent-set-of-relations}) derives only from the $VVP$ Green function analysis and
the requirement of the vanishing of the anomalous form-factors at high energies, one can extract a prediction for the leading term in the expansion, $\mathcal{O}(1/Q^2)$:
the coefficients we obtained in the two form-factors studied here, $F_{\pi\gamma}^V(Q^2)$ and $\mathcal{F}_{\pi^0\gamma\gamma^*}(Q^2)$,
reasonably  agree with the
Brodsky-Lepage result~\cite{BL}, which is preferred with respect to alternative high-energy behaviors from other works~\cite{M, GL}.
\section*{Acknowledgements}
\nin
P.~R. thanks and congratulates Stephan Narison and all his team for the excellent organization and atmosphere of the QCD14 Conference. P.~R. acknowledges
Sergi Gonz\'alez-Sol\'{\i}s for stimulating discussions on this point and also benefited from conversations with Karol Kampf and Marc Knecht on the associated error of
short-distance predictions of resonance Lagrangian couplings and related phenomenology during QCD14. This work is partially funded by the Mexican (Conacyt and DGAPA) and Spanish Governments
and ERDF funds from the European Commission [PAPIIT IN106913, FPA2010-17747, SEV-2012-0249, CSD2007-00042] and the Comunidad de Madrid [HEPHACOS S2009/ESP-1473].








\end{document}